\documentclass{ws-procs961x669}            
\usepackage{enumitem}

\begin{document}
\hyphenation{}

\title{Summary of Parallel Session:\\
``Relativistic and Particle Astrophysics"}

\author{J. \v{R}\'{\i}pa}

\address{
Leung Center for Cosmology and Particle Astrophysics, National Taiwan University\\
No.1, Sec.4, Roosevelt Road, Taipei 10617, Taiwan (R.O.C)\\
E-mail: jripa@ntu.edu.tw
}

\bodymatter
\def\figsubcap#1{\par\noindent\centering\footnotesize(#1)}
 
\section{Summary of the presentations}
The parallel session on ``Relativistic and Particle Astrophysics"
was held on 14th December 2015. Here a short summary of each presentation of this parallel session is given:\\


\noindent
{\bf Sadakazu Haino:} {\em Particle physics and cosmology with Alpha Magnetic Spectrometer}

\begin{itemize}[leftmargin=10pt]
\item[--] The author summarized the history of direct measurements of cosmic-rays in GeV$-$TeV region and the latest results from the four years of data taking with the Alpha Magnetic Spectrometer (AMS) placed on the International Space Station (ISS). The physics implications of the results by AMS were also discussed.

\item[--] The author presented the resent measurements of proton flux\cite{agu15a} and helium flux\cite{agu15b} in primary cosmic rays.

\item[--] Detection of anti-nuclei in cosmic rays would be a strong evidence of primordial anti-matter. Cosmic-ray antiparticles can provide unique opportunity to study fundamental physics such as indirect searches for dark matter and understanding of its nature.

\item[--] The author presented the resent measurements of $\mathrm{e}^+$ fraction and $\mathrm{e}^+$/$\mathrm{e}^-$ fluxes\cite{acc14,agu14} in primary cosmic rays. The AMS measurements of \={p}/p ratio were also discussed.

\item[--] AMS provides the data in an unprecedented precision with a single instrument. 
\end{itemize}


\noindent
{\bf Keiichi Mase:} {\em The dawn of neutrino astrophysics}

\begin{itemize}[leftmargin=10pt]
\item[--] The author described IceCube\cite{ice06}, which has recently observed high energy extraterrestrial neutrinos above a few tens of TeV including several PeV neutrinos\cite{aa13}. Although neutrinos of those energies are predicted to be produced through collisions of hadronic particles accelerated at astronomical sources, they had never been detected before.

\item[--] Neutrino observations in combination with observations by other messengers such as cosmic rays and gamma-rays can provide knowledge about the astrophysical sources that produce them, e.g. AGNs, GRBs, SNe, and cosmogenic neutrinos, which are produced through the interactions of ultra-high-energy cosmic rays (UHECRs) with photons from the cosmic microwave background radiation (CMB). Since neutrinos are weakly interacting particles without electric charge, it is possible to track them back to the source.

\item[--] IceCube also contributes to elementary particle physics by searching for neutrinos from dark matter annihilations and by investigating atmospheric neutrino oscillations\cite{aa15}.

\item[--] The author presented the latest IceCube astrophysical neutrino observations\cite{aa14,kop15}.

\item[--] There are planed extensions of the IceCube experiment: IceCube-Gen2\cite{ice14a} and PINGU\cite{ice14b}.

\end{itemize}


\noindent
{\bf Tsung-Che Liu:} {\em Profile of third flight of the Antarctic Impulsive Transient Antenna (ANITA) in 2015}

\begin{itemize}[leftmargin=10pt]
\item[--] The author gave an overview of the Antarctic Impulsive Transient Antenna (ANITA) experiment\cite{anita09}, which has been designed to study ultra-high-energy cosmic neutrinos(UHECN) and cosmic rays by detecting the radio pulses emitted by their interactions with the Antarctic ice sheet\cite{gor07}.

\item[--] It detects the coherent radio emission (Askaryan radiation\cite{ask62,ask65}) with vertical polarization from neutrino induced particle cascades in ice.

\item[--] ANITA also detects UHECRs by geosynchrotron radio emission with predominant horizontal polarization produced in the atmosphere and reflected by the ice.

\item[--] The third flight of the ANITA was launched on 2015/12/18 and landed on Antarctica after 23-days-long flight. The author introduced the instrument performance during the flight, and showed examples of the collected data.

\item[--] There is a plan for the continuation of this experiment. The ANITA-IV flight is planned in 2016/2017.
\end{itemize}


\noindent
{\bf Chin-Hao Chen:} {\em Searching Ultra-High Energy Cosmic Neutrinos and Flavor Identification with ARA Detector}

\begin{itemize}[leftmargin=10pt]
\item[--] The author presented the current status of searching UHECNs with Askaryan Radio Array (ARA)\cite{all12} in South Pole, and the potential of flavor identification of UHECNs with ARA experiment.

\item[--] If UHEN produces an electromagnetic shower which propagates through a dielectric medium (e.g. ice) it leads to the $\sim20$\,\% excess negative charge. The excess electrons releases the coherent Cerenkov radiation at about 100\,MHz to a few GHz. This is called (Askaryan effect\cite{ask62,ask65}) and ARA uses this effect to detect the highest energetic UHENs in ice. With the attenuation length of $\sim100-1500$\,m, ice is a natural target material for radio detections.

\item[--] Detecting UHECNs\cite{hil06} with energies above $10^{17}$\,eV, or the GZK neutrinos\cite{abb08}, is a fundamental problem in neutrino astronomy. By finding GZK neutrinos, not only the GZK process can be verified, but also provides valuable insights of the ultra-high energy cosmic rays. The author presented the latest results \cite{all15}.

\item[--] Equally important is to identify the flavor of the GZK neutrinos\cite{lai14,aar15} because the flavor composition of these neutrinos provides information about their source.
\end{itemize}


\noindent
{\bf Shih-Hao Wang:} {\em Feasibility of Determining Diffuse Ultra-High Energy Cosmic Neutrino Flavor Ratio through ARA Neutrino Observatory}

\begin{itemize}[leftmargin=10pt]
\item[--] The flavor composition of UHECNs carries precious information about the physical properties of their sources, the nature of neutrino oscillations and possible exotic physics involved during the propagation.

\item[--] The author in his work showed the feasibility to constrain the UHECN flavor ratio by measurement of the angular distribution of the incoming events at a neutrino observatory\cite{wan13}. Simulations were performed assuming the ARA\cite{all12} detector configuration.

\item[--] The author's results assuming the standard oscillation and the neutrino decay scenarios are summarized in terms of the probability of the flavor ratio extraction and resolution as a functions of the number of observed events and the angular resolution of neutrino directions. By fitting the direction distribution of neutrino events, a preliminary constraint on the UHECNs flavor ratio can be set for the planned ARA37 configuration.
\end{itemize}


\noindent
{\bf Guey-Lin Lin:} {\em Thermal transport of dark matter in the Sun}

\begin{itemize}[leftmargin=10pt]
\item[--] The author studied the thermal transport occurring in the system of solar captured dark matter (DM) and explore its impact on the DM indirect search signal\cite{che15}. He focused on the scenario of self-interacting DM (SIDM).

\item[--] The motivation is that Sun is a good target for detecting DM signal through observing neutrinos. DM density in the Sun is expected to be larger than the average DM density in the solar density.

\item[--] The author examined the DM temperature evolution and demonstrated that the DM temperature can be higher than the core temperature of the Sun if the DM-nucleon cross section is sufficiently small such that the energy flow due to DM self-interaction becomes relatively important.

\item[--] He argued that the correct DM temperature should be used for accurately predicting the DM annihilation rate, which is relevant to the DM indirect detection.
\end{itemize}


\noindent
{\bf Jakub \v{R}\'{\i}pa:} {\em Gamma-Ray Bursts and Their Relation to Astroparticle Physics and Cosmology}

\begin{itemize}[leftmargin=10pt]
\item[--] The article gives an overview of gamma-ray bursts (GRBs)\cite{kou12} and their relation to astroparticle physics and cosmology. GRBs are the most powerful explosions in the universe that are characterized by flashes of gamma-rays typically lasting from a fraction of a second to thousands of seconds.

\item[--] There were discovered several correlations in the properties of the GRB emission which makes them a potential tool to constrain cosmological parameters\cite{wan15,ama15} next to the ``traditional" such as SNe Ia or CMB.

\item[--] For example, it was suggested that some of the properties of the prompt and the early optical lightcurves of GRBs could be used as standard candles\cite{panai08}. To confirm that more early optical observations are needed and they can be provided by the Ultra-Fast Flash Observatory pathfinder (UFFO-p)\cite{par13}, which will be able to start optical observations already few seconds after the trigger.

\item[--] GRBs were proposed to be sites for accelerating ultra-high energy cosmic rays and sources of very high energy neutrinos up to $10^{17}\sim10^{19}$\,eV \cite{wax97}. However, the recent results from the IceCube detector suggests that the efficiency of neutrino production may be much lower than predicted\cite{abb12}.
\end{itemize}


\noindent
{\bf Dong-Hoon Kim:} {\em General Relativistic Effects on Pulsar Radiation}

\begin{itemize}[leftmargin=10pt]
\item[--] In this work, the author considers a magnetic dipole model of a pulsar and investigates general relativistic (GR) effects on electromagnetic radiation from the pulsar: the effects on (i) radiation along the magnetic poles and (ii) curvature radiation in the magnetosphere.

\item[--] To determine such effects, the author studies electrodynamics of the pulsar in curved spacetime; by solving Maxwell's equations in curved spacetime\cite{pet74}. In particular, to study electromagnetic radiation near compact objects, e.g. black holes or neutron stars, he applies a ``test-particle" method and a ``semi-relativistic" approximation.

\item[--] The author finds that the major GR effects result from modification of the magnetic field in the strongly curved spacetime of the pulsar. In case (i), modification of the field intensity at the magnetic poles leads to modulation of the frequency of radiation along the poles: e.g. for PSR 1937+214, the field intensity increases by $10.5\pm3.7\,\%$ due to the curved spacetime and this will increase the frequency by the same amount. On the other hand, in case (ii), modification of the field lines in the magnetosphere leads to modulation of the frequency of curvature radiation\cite{kon00}: the radius of curvature of the field lines decreases due to the curved spacetime and this will increase the frequency in an inverse manner.

\item[--] The author also investigates GR effects on polarization of the curvature radiation; by computing the Stokes parameters\cite{gil90} in curved spacetime. The properties of polarization are expected to change due to the curved spacetime: modulation of the radius of curvature leads to modulation of the Stokes parameters, which should cause the ratio between linear and circular polarization to change in curved spacetime. (c.f. Note that Ref.\cite{gil90} is about curvature radiation in ``flat" spacetime, but it is referred to it only for how to compute the Stokes parameters)
\end{itemize}


\noindent
{\bf Hayashinaka Takahiro:} {\em QED Correction of the Radiation from the Magnetors}

\begin{itemize}[leftmargin=10pt]
\item[--] The author investigated the effect of quantum electrodynamics (QED) on the radiation from the magnetars (strongly magnetized neutron stars). The strength of the magnetic field of magnetars is estimated to be higher than $\sim~10^{13}$ Gauss which is the QED scale determined by electron mass. One has to consider the QED effect in such a high energy scales because the photon action or the e.o.m. is corrected by loop effect of electron.

\item[--] He used the Euler-Heisenberg Lagrangian to find the correction for the magnetic dipole radiation which is the leading radiation term due to the lack of electric dipole radiation.  The results were shown for the weak field region ($B\ll10^{13}$ Gauss) and the strong field region ($B\gg10^{13}$ Gauss).

\item[--] The author found the 1-loop QED correction term to the magnetic dipole radiation explicitly and claimed that when the field strength goes extremely large, there would be no correction.
\end{itemize}


\noindent
In addition to the talks in this parallel session, there were two relevant plenary presentations:

\begin{itemlist}[leftmargin=20pt]
 \item {\bf Jiwoo Nam:} {\em Ultra-High Energy Neutrinos and Cosmic Rays via Radio; ANITA, ARA, and TAROGE}
 \item {\bf Henry Wong:} {\em Direct Experimental Searches of Dark Matter}  
\end{itemlist}

\end{document}